# Second-Level Digital Divide: Mapping Differences in People's Online Skills[*]

**Abstract**

Much of the existing approach to the digital divide suffers from an important limitation. It is based on a binary classification of Internet use by only considering whether someone is or is not an Internet user. To remedy this shortcoming, this project looks at the differences in people's level of skill with respect to finding information online. Findings suggest that people search for content in a myriad of ways and there is a large variance in how long people take to find various types of information online. Data are collected to see how user demographics, users' social support networks, people's experience with the medium, and their autonomy of use influence their level of user sophistication.

**Introduction**

Most of the literature on the digital divide focuses on the differences between the haves and have nots of the digital age, that is, the differences among those who have access to the Internet and other communication and information technologies and those who do not, or the differences among those who use these media and those who do not. While most reports identify differences among various segments of the population, over time studies emphasize the increasing diffusion of the medium among the population at large (see Rickert and Sacharow 2000 and Pew 2000a for gender; Austen 2000 for age, NTIA 2000 for race). As more and more people start using the Web for communication and information retrieval, it becomes less and less useful to merely look at demographic differences in who is online when discussing questions of inequality in relation to the Internet. Rather, we need to start looking at differences in how those who are online use the medium. The Web Use Project at Princeton University explores people's skills in locating content online. The study uses rigorous sampling techniques coupled with in-depth interviews and in-person observations of people's online actions to yield data that allow generalizations of findings to more than a small subset of the population.

[*] I would like to thank Paul DiMaggio for his insightful comments on this project. I am also grateful to Edward Freeland, James Chu and Jeremy Davis-Turak for their help with the survey components of the project and to Inna Barmash for her help with coding the data. Generous support from the Markle Foundation is kindly acknowledged. The project has also been supported in part by NSF grant #Z36240, a grant from the Russell Sage Foundation, and through a grant from the Pew Charitable Trusts to the Center for Arts and Cultural Policy Studies, Woodrow Wilson School, Princeton University. A similar version of this paper will also be presented at the Special Interest Group on Information Seeking and Use Research Symposium 2001: Effective Methods for Studying Information Seeking and Use (ASIST, November 3, 2001, Washington, DC).

This paper describes a method that allows us to measure differences in people's skills with respect to Web use and also presents the preliminary findings from interviews with sixty-three participants. First, I briefly discuss the limitations in the current approach to the digital divide and also present the current state of research on Web use for information retrieval. Next, I discuss why the approaches in the existing studies are not always suitable to gaining a refined understanding of the differences in how people locate content online and how this project remedies some of these limitations of existing studies. Then, I describe the project methodology including a discussion of the sampling technique, technical specifications of the project, a detailed description of the two survey instruments, and the in-person observation session of people's online search techniques. Finally, I present preliminary findings based on the first sixty people who participated in the study.

**Approaches to the Digital Divide**

Access to and use of the Internet increased rapidly between 1995 and 2001. The number of Americans online grew from 25 million in 1995 (Pew 1995) to 83 million in 1999 (IntelliQuest 1999) with 55 million Americans using the Internet on a typical day in mid-2000 (Pew 2000b:5). In 1994, just 11 percent of U.S. households had on-line access (NTIA 1995). By December 1998 this figure had grown to 26.2 percent. Less than two years later it stood at 41.5 percent, and well over 50 percent of individuals between the ages of 9 and 49 reported going on-line at home, work, or some other location (NTIA 2000). A November/December 2000 survey by the Pew Internet and American Life project found 58 percent of a national sample online (Horrigan 2000b: 7). Moreover, many more people have Internet *access*, in the sense that they have an available online connection (whether or not they choose to use it) at home, work, school, library or a community center (by the end of 2000, 104 million American adults had access to the medium with [Pew 2001]).

Most reports that explore inequalities with respect to digital technologies concentrate on the divide between those who have or do not have access to the particular medium or those who do or do not use a medium. However, as an increasing portion of the population joins the user universe, it will be important to start differentiating among those who are online without restricting discussions of inequality to non-users alone. The findings of this project suggest that there is considerable variance in the abilities of users to find various types of information online. In so far as accessing information is an important use of the Web and a potential source of inequalities, it is important to explore this aspect of users' online experiences.

A more comprehensive approach to studying the digital divide looks at several dimensions of access and use that may influence people's use of the medium and thus potentially lead to different divides. DiMaggio and Hargittai (2001) suggest that the term "digital inequality" better captures the complexity of inequalities relevant to understanding the differences in access and use of information technologies. Digital inequality considers variation on five dimensions: differences in the technical apparatus people use to access the Internet, location of access (i.e. autonomy of use), the extent of one's social support networks, the types of uses to which one puts the medium, and one's level of skill. The Web Use Project at Princeton University collects measures on all these dimensions for a random sample of Internet users allowing for a more complex exploration of digital inequality.

**Existing Research on Web Use for Information Retrieval**

Scholars from many fields have explored how people use the World Wide Web for information retrieval. Advertising and marketing specialists often refer to users as "consumers" emphasizing their particular interest in people's online actions, namely their shopping behavior (Jarvenpaa, Sirrka and Todd 1996, Bell and Tang 1998). These studies often analyze users' behavior on only one particular site as opposed to exploring users' overall online behavior. Moreover, their sole interest is in how people decide to make online purchases, what influences these decisions, and how much shopping people engage in.

Much work conducted in the human-computer interaction field also tends to concentrate on particulars. Researchers in this area analyze people's use of specific design features and distinct Web site layout (see, for example, the Special Issue on World Wide Web Usability of the <u>International Journal of Human-Computer Studies</u> (1997)). Furthermore, they also look at features of software programs to assess important usability issues (see, for example, Greenberg and Cockburn (1999) for a detailed discussion of the "Back" button on browsers).

Alternatively, computer scientists draw on large-scale aggregate logs about people's use by analyzing all Web activity over a specified period (Catledge and Pitkow 1995, Huberman et al 1998.). An important limitation of many such studies is that they concentrate on the user behavior of a small segment of the population by limiting participants to university faculty and students (e.g. Catledge and Pitkow 1995) or long-term users from the information technology profession (e.g. Choo, Detlor and Turnbull 1999). Admittedly, concentrating on such groups may be legitimate depending on the research questions, however, they limit the extent to which findings can be generalized to a larger segment of the Web user population. In cases where data are derived from larger segments of the online population (e.g. Huberman et al. 1998, Hoelscher 1998, Silverstein et al. 1999, Jansen, Sping and Saracevic 2000), there is no information about specific users and thus it is impossible to make any claims about how attributes of users may be related to their online behavior.

Private research corporations collect data on what sites people visit and how much time they spend on each page (e.g. CyberDialogue and MediaMetrix collect Web behavior information this way). However, such information is proprietary and does not include information on what users are actually looking for (if anything) and whether they are satisfied by the options presented to them on the screen. These data sets also do not contain information on how users perceive what they see and how they make the particular choices in their linking behavior and search strategies.

Researchers in the library and information science community have also conducted numerous studies on people's use of library resources that are often increasingly run on Web-based applications. Abramson looked at how people used the Web at public access computers by recording logs of use via a computer connected to the machines she was observing (1998). However, she only collected information about visited sites and time of day and week without any information about users. Numerous case studies exist on the implementation of specific search programs in libraries (e.g. Payette and Rieger 1997) but these also limit their scope to a distinct user base and Web search protocol or library interface. There are also many studies (Hsieh-Yee 1993, Koenemann and Belkin 1996, Siegfried, Bates and Wilde 1993) that look at searches performed on various information retrieval systems (pre-Web applications as well), however, they focus on the details of query specifics (e.g. number of queries in the data set, session length, query length, use of advanced search functions) without considering information about user demographics or other information retrieval practices of the users.

Closest to the methods presented in this paper are some of the in-person user studies that have been conducted by library and information science researchers. Wang, Hawk and Tenopir (2000) collected data by observing how respondents search for information specified by the research team. Their project generated synchronized video-audio data, which were then analyzed for detailed information about respondents' search techniques. However, as often is the case in such studies, the participants for the study were graduate students and faculty in an information science program. In order to gain a better understanding for how the general population is using the Internet, it is important to include people from beyond the academic community in the studies.

The methods used in the studies cited here provide important information for a baseline understanding of how certain people navigate particular parts of the Web. As Frielander (1997) points out in an editorial about research on digital libraries, referring to the 1970s debates about different methodologies; "quantification was sometimes a good thing and sometimes not; the value of the research resided in the questions it asked and the integrity of the findings, not the methods by which those findings were obtained." There is strength in both quantitative and qualitative approaches to people's use of computers and online resources. Existing studies either limit their scope to specific user populations (e.g. IT professionals or people who go to libraries), do not collect background information about user attributes, or look at use patterns on an aggregate level without collecting data about the specific goals of a Web session.

The Web Use Project adds to the literature on information retrieval in the following important ways: 1. it considers the search patterns of users drawn from the general population instead of solely relying on people in the academic community for data; 2. it collects data not only on users' search activities but also on their use of other media for information retrieval, their demographics and their social support networks; 3. it recognizes that with the Web, searching for information is no longer limited to entering search queries in a search engine, rather, there are numerous ways in which one can go about finding information and these ways may lead to different results and differences in the efficacy of the particular information retrieval technique used. The next section described the methodology in detail followed by some preliminary data analyses.

**Collecting In-Depth Data: Structured Observations and Interviews**

*Sampling*

In order to be able to generalize from the findings, it is important to conduct the study on a random sample of users. The Web Use Project looks at the online use patterns and skills of randomly selected Internet users from the residential communities of an urban (New York City), suburban (Mercer County, NJ) and rural (Pennsylvania) area.[1] A random sample of residential addresses is obtained for each area from Survey Sampling, Inc. and is checked against the National Change of Address Database maintained by the U.S. Postal Service. Potential respondents are first contacted via postal mail. Potential respondents are sent a letter explaining the project and requesting participation with a brochure that presents more details about the study. People are also pointed to http://www.webuse.org on the Web for more information and are given the option of calling/writing to the research center to schedule an appointment. A few days after the letters have been sent, the househoelds are contacted by telephone. The eligible adult (i.e. Internet user adult over 18) with the next nearest birthday is selected in order to randomly sample from within household. People who are identified as Web users are invited to participate in the study.[2] Web users are defined as people who go online at least once every month for more than using email. Although this is a low threshold for including people in the study, it is used to ensure that low frequency users who are nonetheless familiar with the Internet are also included. Asking questions about location of and motivation for use, low frequency of usage coupled with information about the user and his/her online behavior can shed light on some issues of inequality as these factors may influence one's skill sophistication in use of the medium.

People interested in participating are offered $40, which they receive after the observation session. Respondents are asked to come to the research site on the campuses of Princeton University, Pennsylvania State University and New York University respectively.[3] The respondents' email address is recorded and a time for the session is scheduled. Respondents are informed that they will receive a follow-up letter in the mail or an email message (based on their preference) with a reminder and directions to the research site.

---

[1] Trenton is excluded from the Mercer County sample as it is considered an urban area. This paper reports on users from the Mercer County area.

[2] This study only includes adult English-speaking users. Two follow-up studies are already being planned for Spanish-speaking users and high school students.

[3] Respondents in New Jersey and Pennsylvania are offered transportation if they cannot provide their own.

Three days before the study and then the day before, a reminder phone call is placed to the respondent in addition to an email message the day before the study.

*Technical specifications*

Two computers are used for the study to allow for variation in people's computer experiences. For PC users, the computer is a Pentium III running Windows mE connected to the university's network with a 17" monitor. For Macintosh users, there is a G3 iMac (15" monitor) also connected to the university network. The three most popular browsing software applications are all available on both machines. A relatively recent version is used – but not a beta version, to eliminate potential difficulties and ensure that users are not surprised by new features – of Netscape Communicator, Internet Explorer and America Online (the research project has its own AOL account in order not to burden respondents' accounts with the time spent at the session). Availability of both computer types and all three browser applications is important so respondents can use the program with which they are most familiar. The computers are capable of connecting to the Internet on a T1 line. The HyperCam (Hyperionics 2001) software program is used to record the observation sessions on the PCs. This program creates audio-visual files (.avi) of the activity on the screen accompanied by the respondents' comments. HyperCam is able to capture action on the entire screen and thus serves as a very helpful tool for this study. A similar program, SnapZPro (Ambrosia Software 2001), is used on the iMac. Additionally, a program called Don't Panic (Panicware 2001) is used to erase the browser and URL history on each browser program so that each respondent starts out with a clean slate and is not presented with auto-complete options if and when s/he types URLs directly in the location bar, pulls down the location bar menu to select an option or types terms in a form.

The advantages of using such a recording program are manifold. The set up time and program costs are minimal. Although a microphone is used, otherwise the recording is unobtrusive (e.g. it does not require a video camera pointed at the respondent, rather, it records screen shots and verbal communication). The software runs on the machine that is being used so there are no interfering variables. The creation of .avi (audio-visual interleaved) files allows for easy access to data. Storage is also straightforward and allows retrieval of data from numerous locations. Files are stored in a password-protected directory of the university network that is backed-up nightly offering safeguards against data loss. The biggest expense associated with this method (other than compensation to respondents) is the server space required for data storage. A session creates an approximately 500MB-1GB file which can be compressed to 200-600MB.[4] Although this is still a considerable amount of data if one does not have central server space allocation, nonetheless, in the age of read/write CD drives even such file sizes are manageable.

*Interviews*

The in-person sessions start with a twenty-minute interview about basic Web use questions. This interview draws on the Internet Module of the General Social Survey (GSS) 2000. The GSS is conducted every few years on a random sample of the American population with a response rate (70-80 percent) rarely achieved by other surveys. The GSS interviews are conducted face-to-face with people in their homes. For decades, the core section of the questionnaire has been replicated on every survey allowing for time-series analyses about people's political and religious beliefs in addition to a myriad of other attributes. The GSS also contains topical modules that differ from year to year. In Spring 2000, a twelve minute Internet Module was added to ask people about their Internet use at home, work, school and other locations (e.g. libraries).[5] Questions were asked about what online services people use, what type of sites they visit, and how they use the Web for political and cultural activities.

The questionnaire presented to respondents in this study replicates sections of the GSS Internet module in order to allow for comparisons with a larger population of users. The questionnaire is administered verbally to establish a rapport between the researcher and the respondent. Administering the questionnaire right before the observation session proves to be very useful. Because the questions explore many facets of Web use, respondents are prompted to think about numerous details of their Web experiences before sitting down at the computer and embarking on the tasks presented by the researcher. For example, the questionnaire asks users to "think about the last time they viewed content online about a political or social-policy issue, current affair, or a political campaign". Then several questions are posed about this specific experience. By allowing users to reflect on such past experiences, they are given a "warm up" time before their Web use observation begins.

Details are also collected about the types of Web sites users visit in order to know with which types of sites they had had experience prior to the session as opposed to sites they are completely new to during the observation. This is important, as someone who often visits political sites is likely to exhibit different browsing strategies while searching for such Web sources, not necessarily because of a general higher level of skill in this task, but because of prior experience. Information is also collected on people's use of other communication technologies to offer a baseline comparison for how people use the Internet in their day-to-day lives. There are several questions that ask about people's social support networks both online and offline. This information is collected in order to understand how people's networks may influence people's ability to perform online tasks. Again, by asking these questions before the task session, people are prompted to think about the various ways in which they find information including recommendations from friends, family and colleagues.[6]

---

[4] Lower quality recordings may still be adequate for later analysis and will yield much smaller file sizes.
[5] The mean response time of the Internet module was 12:26 and ranged from zero to 45 minutes.

After the observation session (described below), another questionnaire is administered collecting basic demographic information about the respondents (variables such as education, race, family income) in addition to data about the details of their Internet access. Information is asked about the quality of people's connections partly measured by their experiences instead of relying on their technical know-how regarding the modem or connectivity specifications. People's frequency of use is also assessed through several questions, as are their available social support networks. Core demographic information is collected to serve as independent variables in the analysis exploring the determinants of level of sophistication in Web use. Some of these questions are also drawn from the General Social Survey, some are drawn from the HomeNet study at Carnegie Mellon University, while others were added specifically for this study. A long list of computer and Internet related terms are also included on the survey and respondents are asked to rank their understanding of these terms[7]. The goal is to see whether the level of skill measured by analyzing people's actions online correlate with people's scores on these knowledge variables. Because the above method is time- and labor intensive and costly, a longer-term goal of this methodology is to suggest ways in which people's skills can be assessed via survey questionnaires instead of always relying on such detailed studies for such information.

---

[6] See, for example, the following case: After being asked how she would find information about what to do if her wallet was stolen, a respondent offered the following information: "Oh see, I know all about that because on the Internet, a friend sent me a thing saying this is what to do if you ever lose your wallet. I've downloaded it and I keep it." It is important for respondents to signal these distinctions as it is not possible to replicate in the study setting users' email archives and their bookmarks/favorites listings yet it is important for the analysis of users' strategies to know that they do rely on recommendations from people they know for sites they visit.

[7] A list of multiple-choice questions measuring people's actual knowledge of these computer and Internet related terms was added later in the study. The data reported here include the responses to only three multiple-choice knowledge questions that were already administered in the first wave of the study.

*Observation sessions*

The observations are conducted at a university research site. This approach has both advantages and shortcomings. Requesting users to come to a location will affect response rates. It also places people in a location with which they are not familiar and requires them to use a computer that is configured differently from the machine they usually use for browsing. This may influence the results, as certain settings (e.g. the default homepage and bookmarks) are not equivalent to their own. However, this approach controls for quality of Internet connection, and hardware/software differences. It also allows us to concentrate on Web use knowledge in a setting that is equally different and new for all. Moreover, using one computer allows the setup of particular software applications that are required for data recording as described earlier. No default page is set on browsers in order not to influence respondents' initial actions once online. The session is started off by the researcher asking the respondent to recall – if possible – the default homepage on the computer s/he uses the most. The respondent is also asked to comment on how much the browser used in the study in front of him/her resembles the one s/he uses most frequently. That is, the respondent is asked whether s/he has personalized anything on the browser and whether s/he has any bookmarks/favorites set.

Users are given a list of tasks to perform online in order to see how they would find certain information online (see the Appendix for the complete list). These resemble the question on the GSS (Q19) that asks how users would go about finding information about a political candidate. However, instead of the hypothetical question asked on that survey, the researcher is able to watch users go through the process of finding a page and take detailed notes on what they do. Data are also collected about duration on different pages, the amount of time it took to get to a site, and all actions taken to reach the destination. Detailed information is collected and coded on each action including whether the person clicked on a link (and whether that link was an advertisement, an image link, a within page or within frame link), whether the person typed in a URL, what type of search engine the user pulled up, and all of the pages the user consulted before reaching the sought information.

Although a recording of users' visits gives us information about what pages users see (as per the type of data collected by commercial marketing corporations and some studies), it gives us no information on what type of information users were actually looking for and how satisfied they are with what they find. Moreover, most logs of uses do not record information about moves that concern the local cache of the machine. Consequently, such data sets miss information about such details as use of the "Back" button on browsers, which is a considerable problem given that its use comprises up to 30 percent of people's browsing activities (Tauscher and Greenberg 1997) and may be considered a part of one's level of search sophistication.

The task-oriented method is repeated for several types of sites such as arts, current events, volunteer organizations, shopping, music, health-related and job search services. This is important in order to gather information on universal versus topic-specific search strategies. Someone who is universally skilled in finding information may have highly sophisticated skills in locating any type of information whereas topic-specific skills imply that the user has considerably different search skill levels depending on the topic being sought. An example of topic-specific search skills is someone who possesses sophisticated methods for finding Web sites on online music, but has little knowledge of how to arrive at Web pages with scientific information about a health concern. Some of the tasks were familiar activities to some respondents but not to others. However, some of the tasks were new to all respondents which allows for comparison across cases with respect to a formerly uncharted territory (the two tasks that caught most people off guard was one that required searching for a site that compares presidential candidates views with respect to abortion, and the other that asked respondents to look for a page that displays children's art).

During the session, the respondent is encouraged to make comments about his/her actions. S/he is asked if the actions s/he has been asked to perform are ones s/he has performed before. If the respondent gets enthusiastic about showing off a skill or search that s/he has recently engaged in, the researcher encourages her/him to do so even if this action is not directly related to the specific task at hand. Such actions add to the data about users' usual experiences and offer a baseline for what types of online activities users are already familiar with and are likely to perform on a regular basis when not in a study setting. The researcher also asks questions right after the various search sessions. By asking questions about a search session right after it has been conducted and with an exact record of what had happened, it is possible to supplement browser destination data with a better understanding of what drives users' online actions. By talking to people, we learn more about their actions and motivations than if we were simply observing recordings of the pages they visit. In other words, this project is not simply about studying people's sequence of use but also their search strategies, the underlying motivations of their actions, and their levels of satisfaction with their Web experiences. Information is also collected on what alternative

ways – other than the Web – respondents may use to find the requested information. This, again, offers a baseline comparison for how much and for what purposes people tend to use the Internet in their everyday lives.

**Analyzing the Data**

*Tabulating survey responses*

The above methods yield a wealth of data. The verbally administered questionnaires are recorded with pencil on paper and have to be entered into a database. The second survey is presented to the respondents online and thus the responses are automatically coded and can be downloaded in a spreadsheet format.[8] These data are then analyzed via quantitative methods. Each respondent has a corresponding user identification number that is used to match up the three components of the study.

*Classifying search strategies*

The audio-visual (.avi) files generated by the screen capture application are coded while being viewed with a multi-media program such as QuickTime or Windows Media Player. The coding of search strategies is partly based on an exhaustive (albeit not mutually exclusive) list of ways in which one can arrive at a Web site (Hargittai 2000) summarized here:

Previous knowledge of a page:

1. Using the browser's default page.
2. Typing in the URL of a previously visited page, an address that one retained.
3. Using an entry from one's bookmark/favorites list.

No previous knowledge of a page:

1. Try to guess the address of a Web page (e.g. by using the word of the sought information in the URL).
2. Do an open search with the help of a search engine by typing certain terms into a search form (or in the new version of Internet Explorer, which comes with a built-in search engine, typing the term in the location bar itself).
3. Click on a directory category and find links via use of directories and subdirectories.
4. Click on an advertisement.

---

[8] The online survey uses the Princeton University Survey Facility which is an application available to members of the Princeton University community for administering Web surveys (http://www.princeton.edu/~jkchu/Survey/).

5. Recall an address from exposure through another medium (e.g. radio, television, newspaper ad or article, billboard ad, recommendation from a friend or colleague)
6. Click on a link in an email from a friend/colleague/mailing list that contains a site recommendation.
7. Use browser add-ons for link (e.g. NeoPlanet or Alexa).
8. Click on a link from the contents of another page (other then options 3 & 4 above).

In addition to noting the number of the above strategies used by the respondent for the various searches, researchers also take note of other aspects regarding the respondents' online actions and comments about the search. Examples of these include the use of various browser features (the use of the Back button vs. the History file or clicking on links twice vs once, pressing Enter vs clicking on Submit and such details). Moreover, the researcher codes the route taken to the sites by making note of the various sites visited and the types of options that were chosen along the way (e.g. advertisement links, search results, etc.). A note is made about whether the respondent knows how to read URLs (Uniform Resource Locators or Web addresses) and knows how to interpret and manipulate them, whether the respondent knows who is responsible for the various Web sites s/he visits and which parts of the page the respondent looks at. We know such information by having prompted the respondents to talk through their actions. Finally, information about search strategies is matched with how easily and quickly the respondent was able to find the desired information. Again, it is important to note that while someone can be extremely skilled at locating one type of information, they may have fewer skills in locating a different type of information. For this reason, both topic specific and universal search strategies are noted.

**Preliminary Findings**

*The sample*

The findings reported here are based on 63 interviews conducted during Summer 2001 in Mercer County, NJ. Respondents range in age from 18-81 (four percent are in their teens, thirteen percent in their 20s, seven percent in their 30s, eighteen percent in their 40s, nine percent in their 50s, eight percent in their 60s, three percent in their 70s and one person in his 80s). Forty-three percent of this sample is male. Fifty-seven percent of these participants work full time and an additional eight percent work part-time. Their occupations range from real-estate agents, environmental policy analysts, blue-collar workers, office assistants, teachers, service employees and medical professionals in addition to students, unemployed and retired persons. Eighty-eight percent of the respondents are White, six percent are African American and four percent are Asian American. Twenty-six percent live alone, 53 percent live with a spouse, the rest live with roommates or others (parents in most cases). The family income of these people is larger than the national average although it is important to note that Mercer County is one of the highest income counties in the country (moreover because this study excludes inner-city Trenton, the poorest neighborhoods are not in the sample at all).

*Users' skills*

There is a large variance in the amount of time people take to complete all seventeen tasks ranging from twenty minutes to over 100 minutes. Most people are eventually successful in locating most of the requested content although some fail in succeeding with as many as half of the tasks. Respondents were encouraged to continue searching without giving up too easily (a minimum of five minutes was given for each task unless the respondent exhibited frustrations and expressed a need for moving on in which case the researcher read the next task).

Some patterns are emerging with respect to what makes a good searcher, which will be helpful in coming up with ways to educate users about their online actions. Although there are a myriad of ways in which one can improve one's sophistication in finding information online, there are a few basic skills that significantly improve the chances that one will find what one is looking for. Knowing when to use an open-ended search versus browsing through category directories compiled by large sites can help significantly. When looking for something more general such as comparison shopping for a used car, turning to categorized directories and Web guides offered by big sites may be more helpful than doing an open-ended search on a search engine. However, when looking for something very specific such as recipes for lactose-free cooking, a multi-term search will be much more efficient. Knowing some of the intricacies of how to use a search engine can be extremely valuable as well (e.g. use of Boolean operators). People who recognize the value of typing in more than one search term especially in the case of a complex search have a much easier time finding sites that address their queries. Moreover, understanding how search engines rank pages, and being able to read search results (including the URLs (Web addresses) of the results) can be quite valuable.

An interesting finding of this study is the extent to which members of the general user population lack the basics of surfing the Web. A few people barely know what a Back button is and thus have an incredibly hard time moving from screen to screen. Many people know of no search engines and solely rely on functions of their browsers or Internet service providers. Some respondents also had a hard time entering valid search terms. One recurring mistake was entering multiple term queries without any spaces. When asked about this practice, one respondent replied that you are not supposed to use spaces on the Web thus the exclusion of spaces in between search terms.[9] Others' exhibit the exact opposite behavior by typing search terms in the location bar itself. However, given that most browsers now automatically redirect those terms into a search engine, this seems to cause fewer dead-end sessions.

In general, young people (late teens and twenties) have a much easier time getting around online than their older counterparts (whether people in their 30s or 60s). Some of this is clearly based on comfort with the technology they are using and not necessarily based on elaborate techniques they have mastered specifically with respect to the Web. Consequently, although older people (especially in their 60s and older) may be slower on the whole, age does not mean that they do not possess the necessary skills to be efficient searchers.

More findings are forthcoming in later versions of this TPRC Draft paper
and at the talk in Alexandria, VA.

**Please check with author (papers@eszter.com ) for latest version before citing!**

---

[9] Respondents were asked about some of their online actions after the full search session had been completed to make sure that the rest of their search behavior would not be influenced by the researcher's question.

## Appendix – List of Tasks Performed During Observation Session

Participants are given the following instructions during the observation session:

If you can recall, please bring up the page that is usually on your screen when you start using the Web.  That is, the Web site that comes up when you start your Web browser program.

Now please conduct the following tasks.  Please note that there is no right way of doing these tasks. We are interested in seeing how you go about finding the following information online.

1. Where do you look for information about current events?

2. How would you find information about local cultural events in your area such as art shows, musical performances, theatre shows or movies? [wait for R to show preference among these tasks.] More specifically, I would like you to show me how you can find listings for local shows [movies, musical performances, theatre or art shows based on R's preference].

3. Where would you get sport scores online?

4. If you wanted to listen to some music online, where would you look for music that you could listen to right away?

5. Imagine you wanted to use the Web to find out information about a political candidate. What would you do?

6. Now let's assume that you wanted a neutral opinion comparing different presidential candidates' thoughts on a particular issue, say abortion. How would you find a neutral third-party site that compares presidential candidates' views on abortion? [If R challenges the existence of "neutral" sites on the Web then encourage R to focus on finding a site that compares the views of candidates even if the site is not necessarily neutral.]

7. Now think about health information.  Let's assume that you or a friend/family member has been diagnosed with lactose intolerance. Where would you look for information about lactose intolerance?

8. As a follow-up, let's assume that you are having a friend over for a meal and this friend is lactose intolerant.  You need to figure out what type of food you can serve.  How would you find a recipe that is explicitly stated as acceptable for someone with lactose intolerance?

9. Now assume that you were interested in looking for a new job or exploring career opportunities. How would you use the Web for this purpose? [Prompt R to look for something realistic, a job for which they would be qualified/interested in.  If R is retired and would not be looking for a job for themselves, encourage R to look for a job for a friend/relative.]

10. Let's assume that you have decided to buy a car and you would like to use the Web to help you in this. You are interested in getting a used car - say a 1995 Ford Escort - and would like to get some information about its features, its price, and most importantly whether there is one such car available for sale in your area. What would you do? [Be sure R looks for availability of such a car locally.]

11. If you were interested in finding the contact information (phone number, postal address, or possibly email address) of a long lost friend (a childhood friend or classmate), how would you go about it?

12. Let's assume that you have decided you would like to do some volunteer work.  It is up to you what type of volunteer work you would do, or whether you would even have anything specific in mind. How would you use the Web to find information about volunteer opportunities? [Prompt R to look for something realistic that they might actually pursue.  Pursue the interest until R reaches specific contact information.]

13. Let's assume that it is time to do your taxes. Where would you find the necessary forms and get information to help you with this process?  [Ask R to look for the Federal 1040, the actual form.]

14. Imagine you would like to view some art by kids online. How would you go about this?

15. Have you ever viewed the pages of any museums or art galleries online?  Can you show me where you would get such information?

16. Where would you find information about recall notices on children's toys?  [wait..]  Ideally, I would like to see a list of recalls, not just one.

17. Imagine that your wallet was stolen and you wanted to quickly find out what steps you need to take in order to avoid the potential problematic consequences of such theft.  Where would you look for information about how to efficiently react to such a situation?  In other words, you are looking for a resource that explains the three or four essential steps you have to take in such a situation, the phone numbers you have to call and such.

    Also ask: Is identity theft something you would be concerned about?

18.  Please try to recall the last time you purchased something online.  What was it?  Where did you purchase it?  How did you find that site?
    [This is not a task, it is for informational purposes about R's usual surfing activities.]